\documentclass[letterpaper, 10 pt, conference]{ieeeconf}  

\IEEEoverridecommandlockouts                              
\usepackage{graphics} 
\usepackage{epsfig} 
\usepackage{amsmath} 
\usepackage{amssymb}  

\usepackage{url}
\usepackage[ruled, vlined, linesnumbered]{algorithm2e}
\usepackage{verbatim} 
\usepackage{soul, color}
\usepackage{lmodern}
\usepackage{fancyhdr}
\usepackage[utf8]{inputenc}
\usepackage{fourier} 
\usepackage{array}
\usepackage{makecell}
\usepackage{hyperref}

\usepackage[
backend=biber,
style=numeric-comp,
sorting=none
]{biblatex}

\SetNlSty{large}{}{:}

\makeatletter

\newcommand{\Rom}[1]{\expandafter\@slowromancap\romannumeral #1@}
\makeatother

\title{\LARGE \bf
Notes on recasting the ``ATLAS-EXOT-2019-23'' search for pairs of displaced hadronic jets in the ATLAS calorimeter}

\author{\parbox{5 in}{\centering Louie Dartmoor Corpe${^1}$, Thomas Chehab${^2}$, Andreas Goudelis${^1}$\\
\vspace{1cm}
${^1}$\textit{Université Clermont Auvergne, CNRS/IN2P3,  Laboratoire de Physique de Clermont Auvergne, 63000 Clermont-Ferrand, France}
\\
${^2}$\textit{Artemis, Université Côte d’Azur, CNRS, Observatoire Côte d’Azur, BP4229, 06304, Nice Cedex 4, France}}
}

\begin{document}

\maketitle
\begin{abstract}
This note describes the validation of material allowing the reinterpretation of an ATLAS search for decays of pair-produced neutral long-lived particles decaying in the hadronic part of the calorimeter, or at the edge of the electromagnetic calorimeter, using the full Run-2 ATLAS dataset. This reinterpretation material includes an efficiency map linking truth-level kinematic information (decay position, transverse momentum and decay products of the LLPs) to the probability of the reconstructed event being selected in the analysis signal region. In this document we describe the validation procedure, \textit{i.e.} how the map was used to recover the limits presented in the ATLAS publication using events generated with {\tt MadGraph5\_aMC@NLO} and hadronised using {\tt Pythia8}, and we identify some limitations of this approach. We moreover comment upon issues concerning the validation procedure itself, in particular with regards to whether or not the information included in the existing, published material allows for an external user to test recasting methods.
\end{abstract}

\section{\textbf{Introduction}}
\label{sec:intro}

The decay of a long-lived particle (LLP) produced in proton–proton collisions in the ATLAS detector could give rise to one of a variety of highly unconventional signatures, depending on the properties of the LLP and the region of the detector where the decay took place. Searches for promptly-decaying particles often have very low sensitivity to such signatures\footnote{For a recent attempt to employ prompt searches in order to constrain LLPs \textit{cf e.g.} Ref.~\cite{Corpe:2024ntq}.} and, hence, the study of LLPs requires dedicated analyses.

\begin{figure}[!ht]
	\begin{center}
		\includegraphics[scale=0.30]{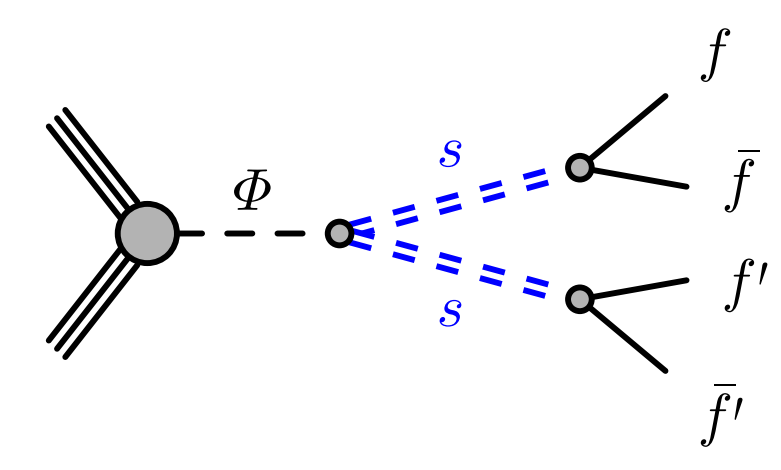}
		\caption{\normalsize{Schematic diagram of the process $\Phi \xrightarrow[]{}ss\xrightarrow[]{} f \overline{f} f' \overline{f'}$ decay used as a benchmark model. The LLPs couple to SM fermions in a Yukawa-like manner, via their mixing with $\Phi$, and therefore decay primarily to heavy quarks \cite{ATLAS:2022zhj}}. \label{diag_toy}}
	\end{center}
\end{figure}

In particular, Ref.~\cite{ATLAS:2022zhj} (EXOT-2019-23) presents a search sensitive to neutral LLPs decaying in the calorimeters of the ATLAS detector. The analysis is capable of probing lifetime values ranging between a few centimetres up to a few tens of meters. This search set limits on a hidden-sector benchmark model in which a pair of long-lived scalars $s$ with masses between 5 GeV and 475 GeV is produced through a scalar mediator $\Phi$ with a mass ranging from 60 GeV to 1 TeV, as represented in Fig.~\ref{diag_toy}.

Depending on their nature, the fermions that come from the LLP decay may result in jets that can appear far from the interaction point (``displaced jets/vertices''). Ref.~\cite{ATLAS:2022zhj} focuses on decays taking place in the calorimeters: either towards the edge of the electromagnetic calorimeter (ECal) or in the hadronic calorimeter (HCal).  In such cases, the LLP decay products can be reconstructed as a single jet which is typically trackless, narrower and with a higher proportion of its energy deposited in the HCal than in the ECal compared with SM-induced hadrons. Pair-produced LLPs are considered, thus this analysis requires two such non-standard jets, with two selections targeting different LLP kinematic regimes. One is optimized for models with $m_\phi$ > 200 GeV (High-$E_\textrm{T}$) and one for models with $m_\phi$ < 200 GeV (Low-$E_\textrm{T}$), where $E_\textrm{T}$ stands for the jet transverse energy. Dedicated techniques, heavily reliant on machine learning and the detailed response of the detector, were developed for the reconstruction of displaced jets produced by LLPs that decay hadronically in the ATLAS HCal.
The resulting search presented selection efficiencies for various parameter choices in the underlying benchmark model.

In the original study, the background was evaluated using a data-driven ABCD method. In a plane of two uncorrelated variables, four regions $A,B,C,D$ were defined. For the assumed underlying model, the signal is chiefly found in region $A$ while the background contribution in that region should be obtained from a simple relation ($A=B \times C/D$) of the yields in the other regions. Since no excess is seen, a simultaneous signal-plus-background fit is used to produce upper limits on the LLP pair-production cross-section times branching fraction.

But what if one wanted to evaluate the impact of this analysis on a model with slightly different kinematics? Re-running the entire analysis or manually reproducing the cutflow is impracticable for an external user, given the highly detector-dependent selection and heavy use of machine-learning methods. However, the HEPData analysis record~\cite{HEPData} \textit{did} provide a new form of efficiency map linking the detector response and selection probability to truth-level information about the LLPs, folding in detector effects. 

In this note, we seek to validate this efficiency map. The code we used to generate the benchmark events, read the map, evaluate the efficiency as a function of lifetime, and set limits, is available on github~\cite{thomascode}.

\section{\textbf{Re-interpretation material}}

The efficiency map is provided in \texttt{.yaml} format in the HEPData record for the search~\cite{HEPData}. This record also provides some template code (under {\tt Resources > Python file}) to read the map and return per-event selection probabilities. The map takes as input truth-level kinematic information about the LLPs and returns a probability that the event would be selected in region A for the analysis, at the detector level. It can in some respects be thought of as a ``folding matrix'' and relies on the fact that the detector should not care what the details of the model were, but should have the same probability of selecting a neutral LLP decay in any given section of the detector, for given transverse momentum $p_\textrm{T}$ and final state (codified through the decay product particles' PDGID). 
The map is provided in two variants, one for the High-$E_\textrm{T}$ selection and one for the Low-$E_\textrm{T}$ selection. The corresponding efficiencies (to be defined shortly afterwards) are shown in Fig.~\ref{Map}. 

It should be noted that the maps implicitly assume that a new model being evaluated would pass two important selection criteria: first, that the sum of missing hadronic energy in the event should be less than 0.6 times the total hadronic energy. Secondly, that the jets in the event must be ``trackless'' (in practice this means that they should not have tracks within a cone of radius 0.2). We will further comment on these criteria in what follows.

\subsection{\textbf{Structure}}

\begin{figure}[!ht]
	\begin{center}
		\includegraphics[width=0.4\textwidth]{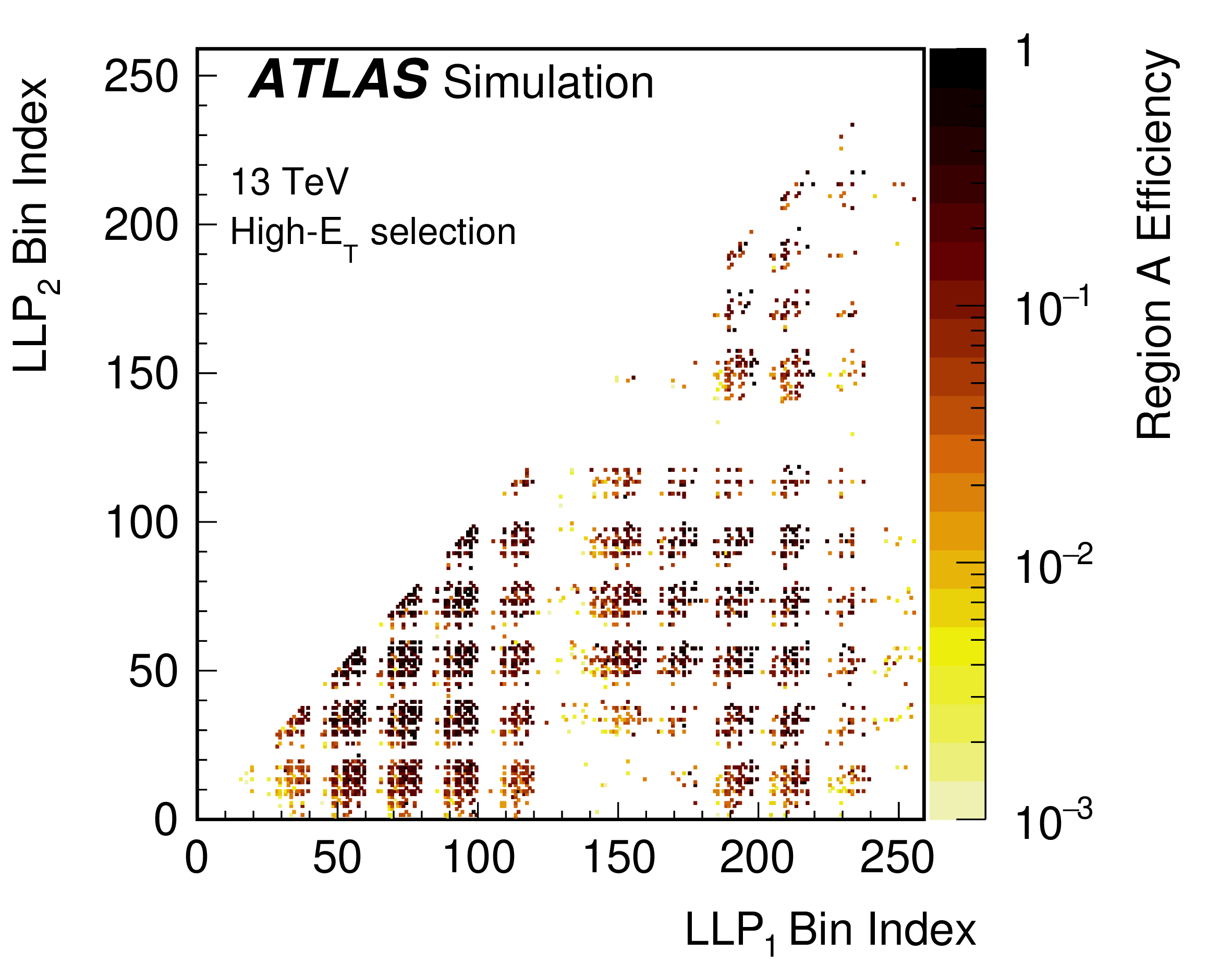}
  		\includegraphics[width=0.4\textwidth]{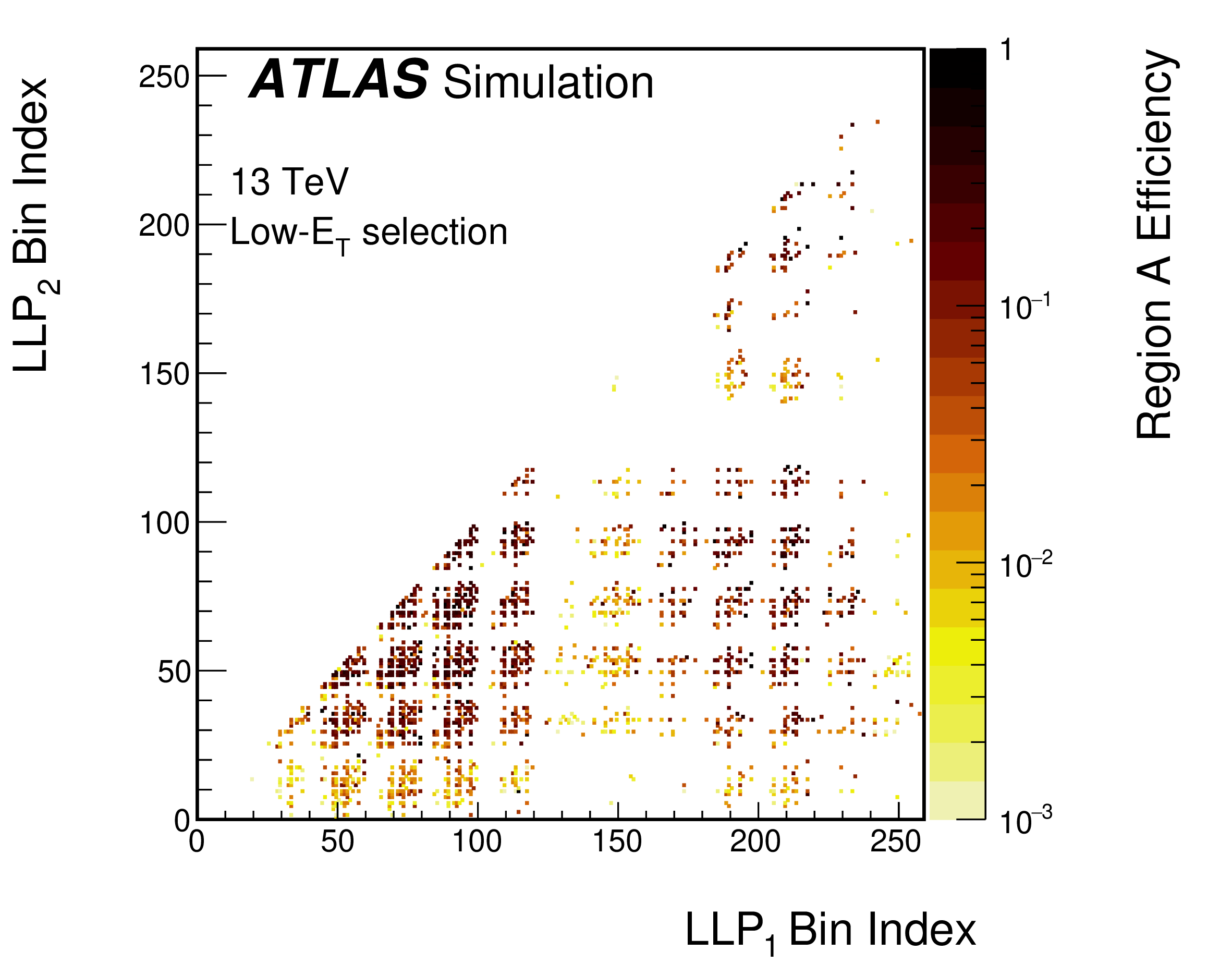}
		\caption{\normalsize{The two re-interpretation maps provided by the ATLAS analysis \cite{ATLAS:2022zhj}. The High-$E_\textrm{T}$ map is shown on top, the Low-$E_\textrm{T}$ map is shown at the bottom. The definition of the ``Bin Index'' is given in the main body of the text.} \label{Map}}
	\end{center}
\end{figure}

Each bin of the map outputs a number between 0 and 1, which corresponds to the probability of an event being selected in region A of the search for a pair of LLPs given:
\begin{itemize}
 \item their decay positions (in the transverse direction $L_{xy}$ for the barrel or the longitudinal direction $L_z$ for the endcap, depending on the pseudorapidity $\eta$);
 \item the LLPs' transverse momenta;
 \item the decay products' PDGIDs.
\end{itemize}
The sum of the output across a sample is an approximation of the total number of events passing the selection. To obtain the efficiency of a given sample, one should sum the output values for each event as extracted from the map and divide by the total number of events in the sample (adequately weighted if appropriate). As we can see in Fig.~\ref{Map}, the map is symmetric between the two LLPs, The choice of LLP 1 and 2 is arbitrary. This map folds in all detector and analysis effects, including triggering and machine-learning discriminants.

The LLP transverse momentum ($p_\textrm{T}$) is binned in the range of $\left[0, 50 , 100, 200 , 400, 1600\right]$ GeV (5 bins), the LLP transverse decay position $L_{xy}$ is binned in $\left[0, 1.5, 2, 2.5, 3, 3.5, 3.9, \infty\right]$m (7 bins), and the longitudinal decay position $L_z$ in $\left[0, 3.6, 4.2, 4.8, 5.5, 6, \infty\right]$m (13 bins). These decay position bins are concatenated, with the $L_{xy}$ bins used for decays in the barrel ($|\eta|<1.5 $), and $L_{z}$ bins used for decays in the endcap ($|\eta|>1.5 $).

Note that in the forward region, the map does not assign a probability of zero to events involving one or more LLPs outside the acceptance region ($|\eta|>2.5) $: the proportion of outside-of-acceptance events is, in a sense, folded into the probabilities returned by the map. Nevertheless, if the $\eta$-distribution of LLPs in any new model varies significantly from that of the training samples, this could have an impact. Such effects could, however, be corrected post-hoc in any re-interpretation study, \textit{cf} the discussion in Section \ref{sec:lessonmap}.

Four decay modes are considered by the map, namely pairs of $c$, $b$, $t$, $\tau$ in bins [0,1,2,3] respectively. 
These were, indeed, the principle decay products predicted by the benchmark models used in the original analysis. We expect, although we wish to be clear on the fact that we have not explicitly checked this statement, that decays into $u, d, s$ quarks or gluons would have a similar efficiency as decays into $c$-quarks. Other decay channels (for example into electrons and muons) should be assigned a probability of 0. The final efficiency is presented as a function named "Bin Index'', defined as: 
\begin{align*}
     \text{Bin Index} = & \text{decay position bin index}   \times (\text{number of $p_\textrm{T}$ bins} \\ & \times \text{number of decay type bins}) \\ & + \text{$p_\textrm{T}$ bin index} \times {(\text{number of decay type bin})} \\ & + \text{decay type bin index} 
\end{align*}
Note that the bin index does not need to be calculated manually: the helper code published on the EXOT-2019-23 HEPData page includes a function to do so. 

From the four regions previously introduced (A,B,C,D), only one is approximated by the map, namely region A.
This constitutes another assumption to bear in mind in future re-interpretation attempts: one would need to assume that most signal events fall within Region A rather than B, C or D.

\subsection{\textbf{Map uncertainties}}
\label{uncert}

The description of the maps provided in Ref.~\cite{ATLAS:2022zhj} includes an estimate of their associated uncertainty, which was derived from internal closure tests on the benchmark samples (without lifetime extrapolation). The following estimates are quoted in the captions accompanying the maps:

\begin{itemize}
    \item For the High-$E_\textrm{T}$ map, for overall efficiencies above 0.5\%, the results are typically accurate at the 25\% level, but below this threshold the efficiency can be overestimated and therefore these values should not be used for re-interpretation.
    \item For the Low-$E_\textrm{T}$ map, for efficiencies above 0.15\%, results are typically accurate at a level of 33\%. Below this value the efficiency is typically accurate up to factor of 3.
     \item The map is less accurate for the Low-$E_\textrm{T}$ samples because the efficiency values tend toward zero.
\end{itemize}
These uncertainties are clearly ad-hoc and this approach is obviously far from ideal. As we will see, it is sometimes insufficient in regions away from the target lifetime range of the analysis.

\section{\textbf{Computational framework for validation}}

The code which we have used to validate the map (and which involves generating events, drawing random lifetimes, evaluating the efficiency and calculating new limits) is available at Ref.~\cite{thomascode}.
The instructions to install the dependencies for this code can be found in:
\texttt{CalRatioDisplacedJet/READ\_ME}.

\subsection{Benchmark model}

The original ATLAS analysis presented in Ref.~\cite{ATLAS:2022zhj} was performed using the so-called ``Hidden Abelian Higgs Model'' (HAHM) \cite{Gopalakrishna:2008dv,Wells:2008xg} as a benchmark model. In this scenario, the role of the mediator $\Phi$ (\textit{cf} Fig.~\ref{diag_toy}) is played by a Higgs-like boson, whereas that of the LLP by a ``dark Higgs'' which is responsible for the breaking of a hidden $U(1)_X$ gauge symmetry. The masses of both particles are taken to vary as described in Section \ref{sec:intro}. Since here we are mainly interested in recovering the ATLAS results, we will be using the same model for event generation. A {\tt FeynRules} \cite{Alloul:2013bka} implementation of the HAHM was presented in \cite{Curtin:2014cca}. Note that in this implementation the mediator $\Phi$ is produced primarily through gluon fusion, which is described by an effective $ggh$ vertex.

\subsection{Event generation, efficiency evaluation and limit-setting}

In order to generate our event samples we used {\tt MadGraph5$\_$aMC@NLO} (MG) \cite{Alwall:2014hca} employing the $NNPDF2.3Lo$ set of Parton Distribution Functions. Showering and hadronisation were simulated with {\tt PYTHIA 8} \cite{Bierlich:2022pfr} with all settings chosen at their default values.

\begin{figure}[!ht]
    \begin{center}
       \includegraphics[scale=0.55]{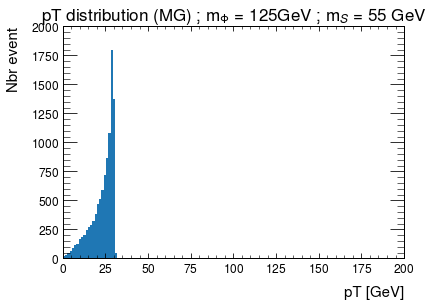} 
      \includegraphics[scale=0.55]{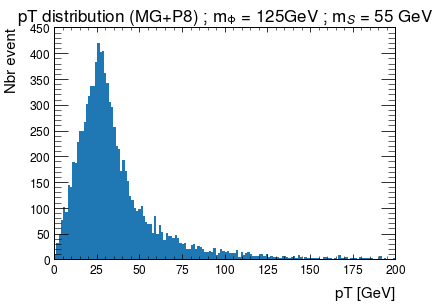} 
      \caption{\normalsize{Comparison of distribution for the transverse momenta without and with the hadronisation for a signal with low mediator mass.} \label{comparison_distrib_pT_low}}
  \end{center}
\end{figure}

\begin{figure}[!ht]
    \begin{center}
      \includegraphics[scale=0.55]{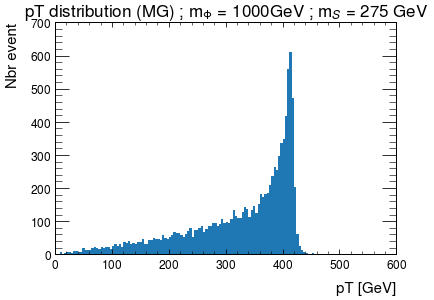} 
      \includegraphics[scale=0.55]{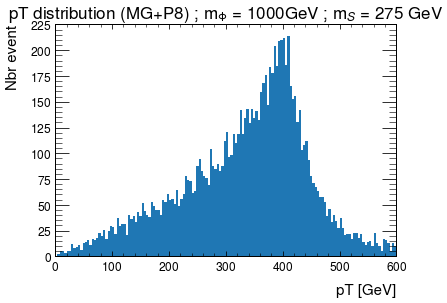}

      \caption{\normalsize{Comparison of distribution for the transverse momenta without and with the hadronisation for a signal with high mediator mass.} \label{comparison_distrib_pT}}
  \end{center}
\end{figure}

In Figs.~\ref{comparison_distrib_pT_low} and~\ref{comparison_distrib_pT} we show, for comparison, the jet $p_\textrm{T}$ distributions without (top panel) and including (bottom panel) hadronisation effects for two example sets of mediator and LLP masses. Although we will further comment on this point in what follows it is intuitively clear that, generically, running {\tt PYTHIA 8} is important when evaluating the efficiency using the map, since the kinematic spectra which are predicted (and which are the inputs of the map) may be modified by hadronisation effects.

All in all, we generated samples for six different benchmark points, which are presented in Table~\ref{massDHetLLP}. 
Note that in principle, once all couplings and masses are fixed, the LLP lifetime is a \textit{predicted} quantity in the HAHM model. In practice, in the analysis performed in Ref.~\cite{ATLAS:2022zhj} the mean proper lifetime of the LLP scalars is treated as an additional free parameter regardless of the underlying actual free parameter value choices, and the event-per-event LLP rest frame lifetime is drawn randomly from an exponentially decaying distribution.

All the parameters and run cards that were used in our analysis can be found in Ref.~\cite{thomascode} in  the script:
\texttt{CalRatioDisplacedJet/Writing\_Scripts\_MG+P8.py}.
\\
\\
Once the events are generated, we calculate the decay position, and use the truth-level information to calculate the ``Bin Index'' for each LLP. We can then extract from the maps the per-event probability that the event would be selected in Region A. By summing the probabilities for each event and dividing by the total number of events, we obtain the sample's selection efficiency.

\begin{table}[!ht]
\centering
\begin{tabular}{|c|*{5}{c|}}\hline
\makebox[11em]{Mass of $\phi$ in GeV}&\makebox[11em]{Mass of the LLP in GeV}&\makebox[6em]{Nbr events}\\\hline\hline
1000   & 275 & 10000 \\\hline
600  & 150 &  10000 \\\hline
400  & 100 & 10000 \\\hline
200  & 50 & 50000 \\\hline
125  & 55 &  50000\\\hline
60   & 5 &  50000\\\hline
\end{tabular}
\caption{\normalsize{Mass of the Dark Higgs and of the LLP generated with MG and the number of events for each sample. \label{massDHetLLP}}}
\end{table}

The code also produces limits: for this, a single-bin fit was performed for the observed and expected number of events in Region A from the original paper, with a dummy signal strength for an arbitrary number of predicted signal events.
The limit on the signal strength is then rescaled by the aforementioned sample efficiency, for each lifetime value, to produce limits on the model cross-section times branching fraction as a function of $c\tau$ ($c$ times the mean proper lifetime of the LLP).

\section{\textbf{Validation results}}

\subsection{Efficiency results}

\begin{figure}[!ht]
    \begin{center}
      \includegraphics[scale=0.45]{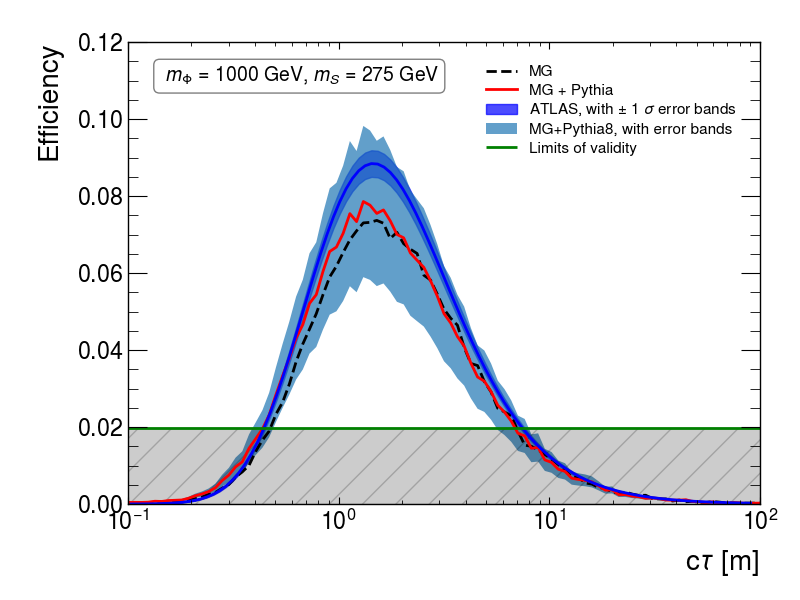} 
      \includegraphics[scale=0.45]{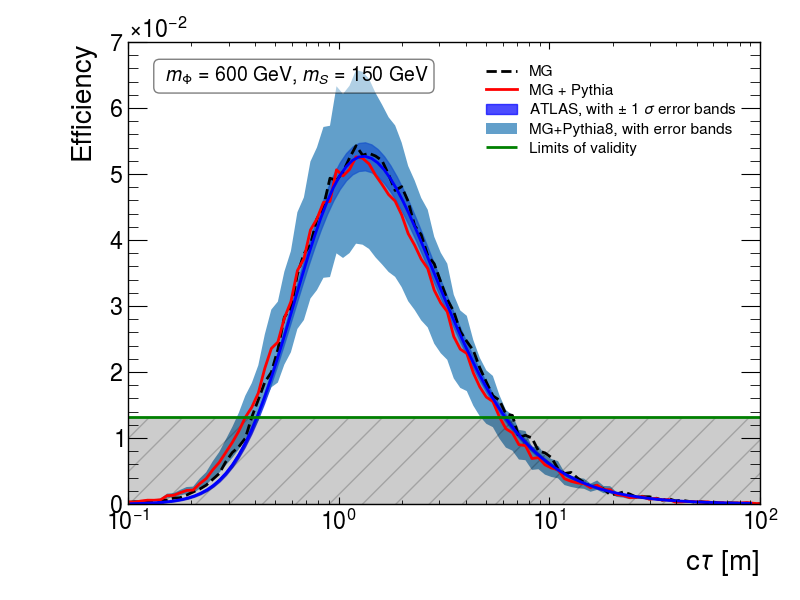} 
      \includegraphics[scale=0.45]{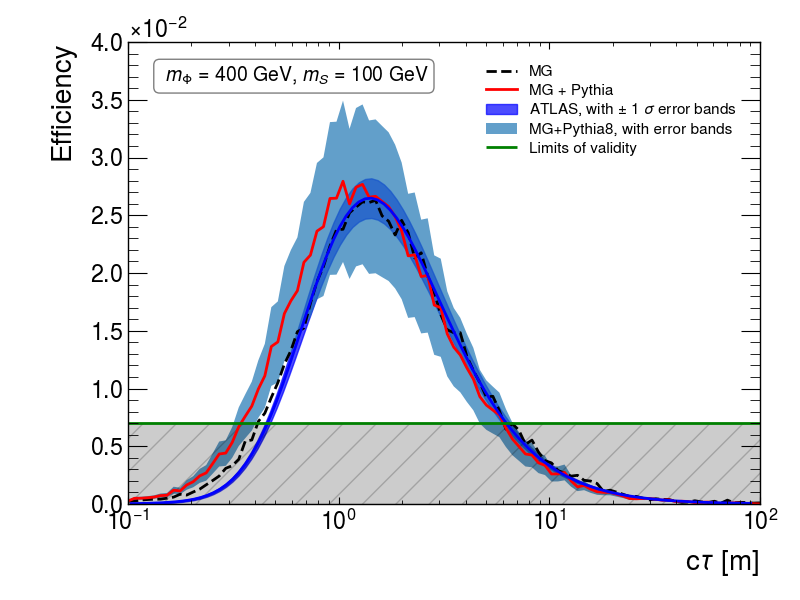} 
      \caption{\normalsize{Efficiencies obtained for  samples with m$_{\phi}$ = 1000, 600, 400 GeV and m$_s$ = 275, 150, 100 GeV using the High-$E_\textrm{T}$ selection, compared to the origin ATLAS analysis.} \label{comparison_1000}}
  \end{center}
\end{figure}

\begin{figure}[!ht]
    \begin{center}
      \includegraphics[scale=0.45]{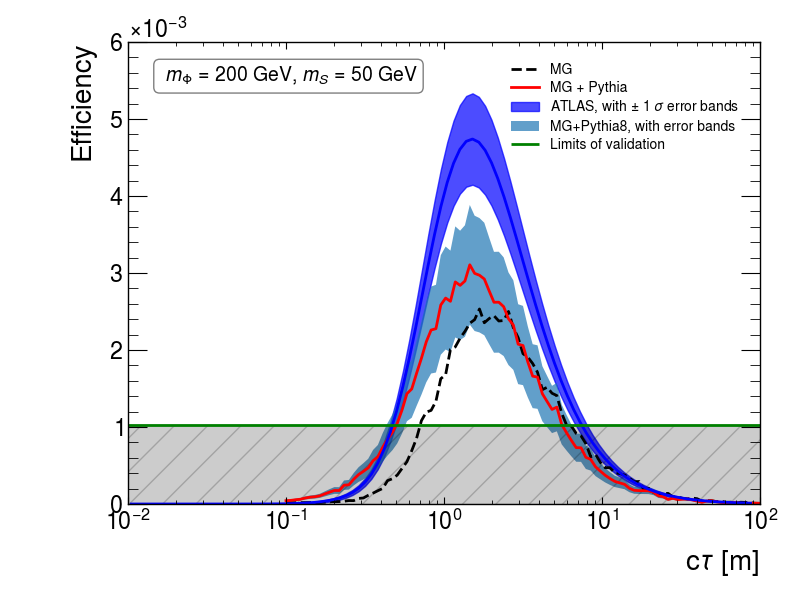} 
      \includegraphics[scale=0.45]{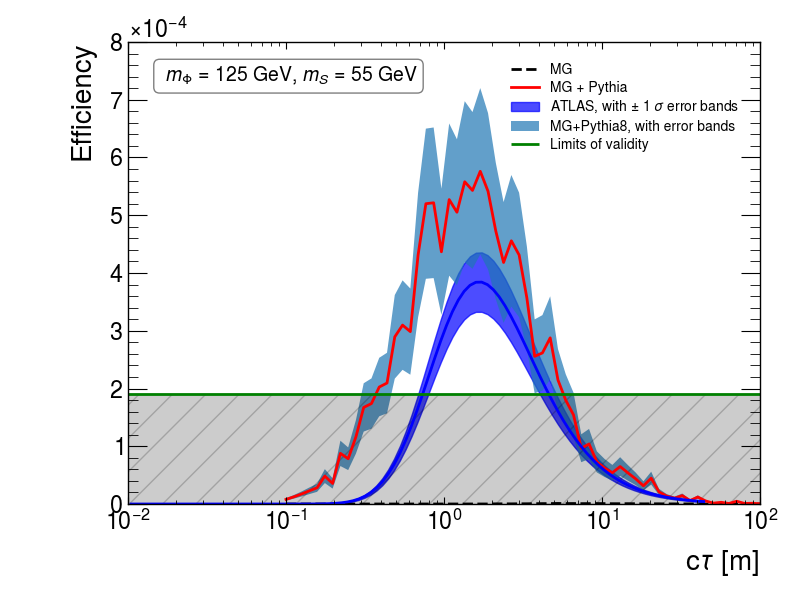} 
      \caption{\normalsize{Efficiencies obtained for samples m$_{\phi}$ = 200, 125 GeV and m$_s$ = 50, 55 GeV using the Low-$E_\textrm{T}$ selection, compared to the original ATLAS analysis.} \label{comparison_200}}
  \end{center}
\end{figure}

In order to validate both the map and the procedure, we compare the efficiency values obtained by passing our generated event samples through the map with the results presented in Ref.~\cite{ATLAS:2022zhj} by the ATLAS collaboration.
The comparison is shown in Figs.~\ref{comparison_1000} and~\ref{comparison_200}.
The ATLAS results are shown in dark blue with the uncertainty bands taken from the original search. The output obtained using the efficiency maps, in turn, is shown in dashed black for MG only and in solid red for MG followed by hadronisation through Pythia8.
This last line also includes the map uncertainties and validity limits described in Sec.~\ref{uncert}, depicted by the light blue shading and green line with hatched region, respectively.

We can see that for the High-$E_\textrm{T}$ benchmarks (Fig.~\ref{comparison_1000}), the results obtained through the map reproduce fairly well the ones obtained through a full-blown analysis, once all uncertainties are properly taken into account. Interestingly, moreover, in this case the results from MG-only and from MG+Pythia agree fairly well: the differences in kinematics due to hadronisation only have a small effect on the final outcome.

On the other hand, the entire procedure becomes less reliable once we switch to the low-$E_\textrm{T}$ samples (Fig.~\ref{comparison_200}). Note also that, according to our findings, showering/hadronisation effects become crucial for the low-mass benchmarks. This can be understood physically because without the recoils introduced by hadronisation, the $p_\textrm{T}$ thresholds for the trigger and analysis are impossible to achieve for the LLPs given the mass of the parent mediator. Besides, the values obtained for the sample with $m_\phi$ = 60 GeV and $m_s$ = 5 GeV are too low to yield usable results. 
Nevertheless, within the region of validity of the maps, the results do agree within order of magnitude.

\subsection{Cross section limits results}

\begin{figure}[!ht]
    \begin{center}
      \includegraphics[scale=0.45]{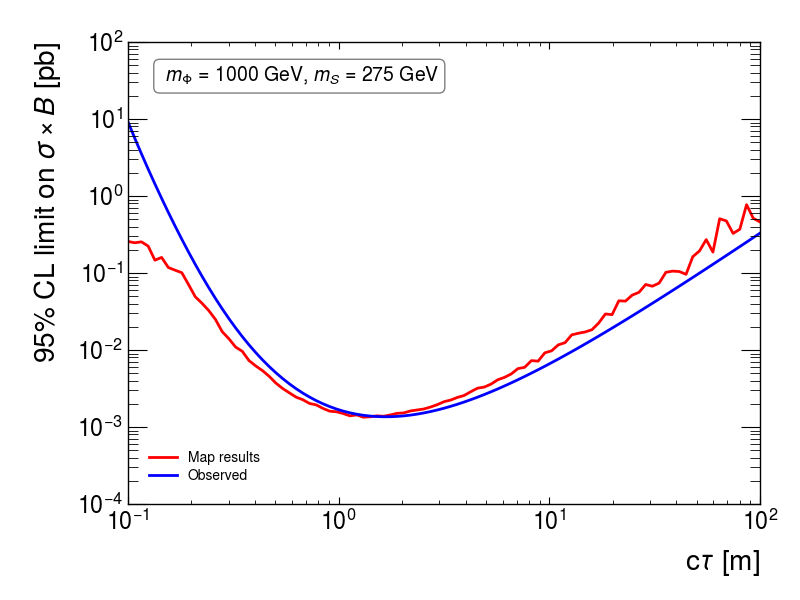} 
      \includegraphics[scale=0.45]{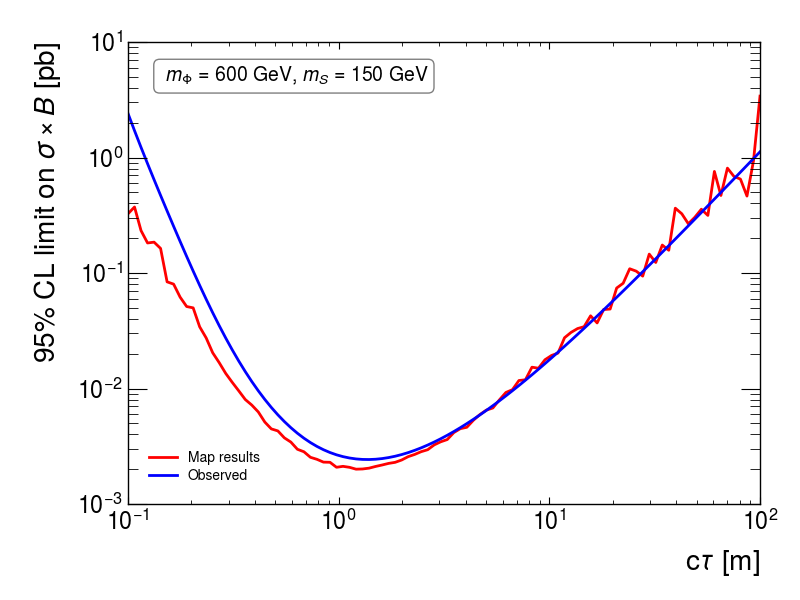}
      \includegraphics[scale=0.45]{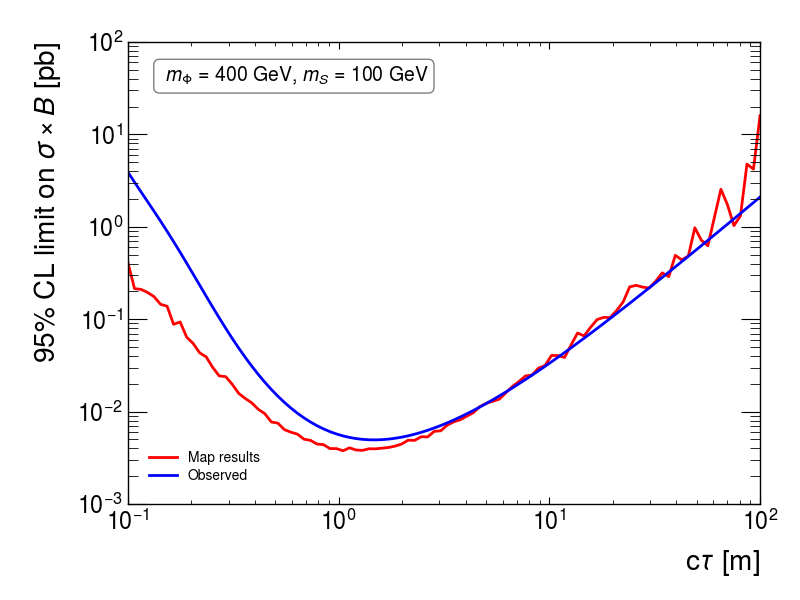} 
      \caption{\normalsize{Limits for the samples m$_{\phi}$ = 1000, 600, 400 GeV and m$_s$ = 275, 150, 100 GeV using the High-$E_\textrm{T}$ selection, compared to the original ATLAS results.} \label{limits_High}}
  \end{center}
\end{figure}

\begin{figure}[!ht]
    \begin{center}
      \includegraphics[scale=0.45]{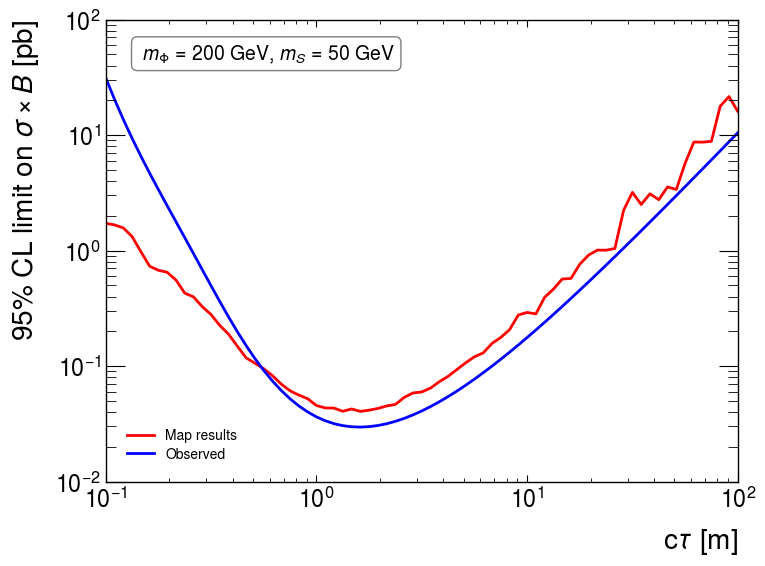} 
      \includegraphics[scale=0.45]{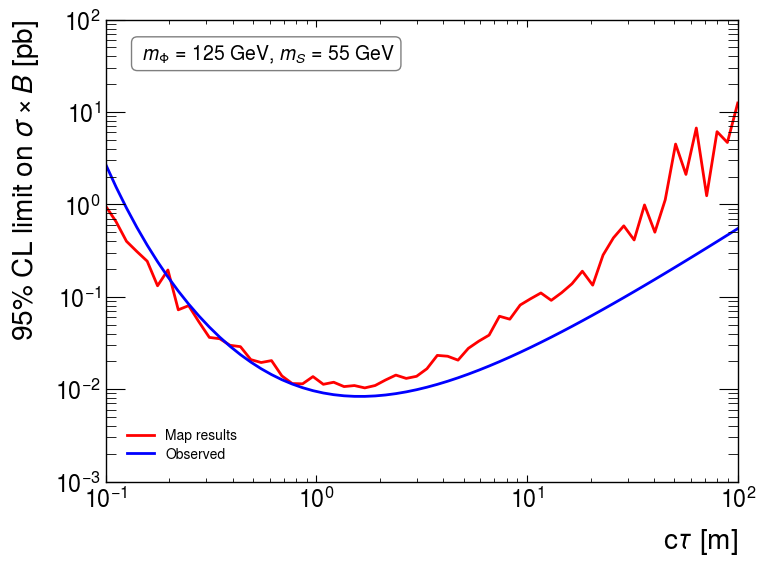}
      \caption{\normalsize{Limits for the samples m$_{\phi}$ = 200, 125 GeV; m$_s$ = 100, 55 GeV using the Low-$E_\textrm{T}$ selection, compared to the original ATLAS results.} \label{limits_Low}}
  \end{center}
\end{figure}

As for the efficiencies, in order to validate the procedure we have plotted the limits that we can set on the cross section by passing our generated event samples through the map and by comparing the constraints to those presented in Ref.~\cite{ATLAS:2022zhj} by the ATLAS collaboration. Our results are shown in Figs.~\ref{limits_High} and~\ref{limits_Low}

As before, we can see that the limits set through the map on High-$E_\textrm{T}$ benchmarks (Fig.~\ref{limits_High}) fit pretty well with the ones obtained through a full-blown analysis, apart from the low lifetimes values. 

\section{\textbf{Conclusions from the validation procedure}}

All in all, our validation procedure allows us to draw several useful conclusions concerning on one hand the map itself, as well as on the process of utilizing it as an external user. In this Section we swiftly present some comments which we find to be the most relevant.

\subsection{\textbf{Comments on the map}}\label{sec:lessonmap}

Let us first make a few remarks concerning the re-interpretation map, the validity of some of the approximations it relies on and the results it allows a user to obtain.
\begin{itemize}
	\item The full ATLAS search used a simultaneous signal-plus-background fit over all ABCD regions, whereas the present treatment only relied on region A results: according to our findings, this is a valid approximation. However, for a new model this conclusion may possibly break down if a large fraction of the signal happens to end up in another region. This seems likely to be a small effect, but it's not immediately clear how this hypothesis can be tested (other than by repeating the entire ATLAS analysis with a radically different model). 
	\item The limits set by relying on the map are fairly realistic for large enough lifetime values but become too aggressive at lower values. One could argue that the map is only safe to use for $c\tau$ values above about $50$~cm.
    \item The map does not account for the fact that some models may have different $\eta$ distributions leading to more particles being lost in the region lying beyond the detector acceptance. The map implicitly assumes that the $\eta$ distribution of the LLPs would resemble the one obtained in a HAHM-like model. This could potentially lead to an over-estimate of the efficiency, so care should be taken. An ad-hoc fix could be to compare the fraction of LLPs in the high-$\eta$ region compared to the HAHM model, and derive a scaling factor from that.
    \item The map similarly incorporates selections related with tracklessness and with the fraction of missing hadronic energy in the event. Implicitly, it assumes that a new model would pass these selections, so again care should be taken to check this for any new model being re-interpreted.
	\item Hadronisation effects are essential during event generation, especially for low-mass mediators. In Fig,~\ref{comparison_distrib_pT_low} we can see that if hadronisation is ignored, the transverse momentum distribution collapses close to $m_s/2$ ($\approx$ 27 GeV), which is close to the experimental limit for a jet to pass selection cuts. 
    \item The obtained results become less precise as the mass of the particles involved decreases. Moreover, in order to avoid wildly oscillating results, a larger number of events needs to be generated.  
    \item The evaluation of the map is relatively slow, and required a custom code, such as the one we presented here. It would be better if it could be provided in a ``standard'' format which could easily be read by existing recasting frameworks.
\end{itemize}

\subsection{\textbf{Comments on event generation from an external user's perspective}}

Lastly, let us point out a few issues encountered during the validation, which hindered a comparison of the  results obtained through the map with the results obtained by the ATLAS collaboration:
\begin{itemize}
	\item The benchmark model employed in the ATLAS analysis is not explained adequately and essentially no links to the employed computational infrastructure ({\tt FeynRules} implementation \emph{etc.}) are provided.
	\item Some key issues concerning the usage of the model are not stated explicitly. In particular, as we already alluded to, even though the LLP mean proper lifetime is, in principle, a quantity which can be \textit{predicted} in the HAHM, in the actual analysis it was treated as a free parameter \textit{on top} of the actual free parameters of the model.
	\item Lastly, several other external or internal parameters of the run are changed by hand throughout the original analysis and not documented (masses, width calculation, cuts).
\end{itemize}


In summary, it would be highly desirable (in reality, absolutely essential) for experimental collaborations to provide a much more detailed presentation of the Monte Carlo event generation process, in order to facilitate recasting attempts.

\section{\textbf{Summary}}

This note briefly described a powerful tool for data reinterpretation contained in ATLAS-EXOT-2019-23: a map that only needs a few truth-level parameters as input to approximate the probability that an event would be selected in the signal region. It is an object in six dimensions where each bin outputs a probability to select an event in the analysis. In order to validate this tool, we placed ourselves in the exact same theoretical and computational framework as the one employed by ATLAS and generated event samples, subsequently comparing the results with the ATLAS public data. 

According to our findings, the recasting procedure manages to approximate the published results in a satisfactory manner, in particular for high-$E_\textrm{T}$ benchmarks, while remaining relatively simple for an external user to utilise. Lastly, as a contribution to the ongoing discussion concerning recasting, data re-intepretation and the presentation of experimental results, we summarised the difficulties encountered during our study, again \textit{placing ourselves at the standpoint of an external user} that aims to re-interpret published data.

\addtolength{\textheight}{-0cm}   




\section*{\textbf{Acknowledgments}}

The authors are grateful towards D. Curtin for sharing the {\tt FeynRules} model files used in the original ATLAS analysis.

\printbibliography

@article{ATLAS:2022zhj,
    author = "Aad, Georges and others",
    collaboration = "ATLAS",
    title = "{Search for neutral long-lived particles in $pp$ collisions at $ \sqrt{s} $ = 13 TeV that decay into displaced hadronic jets in the ATLAS calorimeter}",
    eprint = "2203.01009",
    archivePrefix = "arXiv",
    primaryClass = "hep-ex",
    reportNumber = "CERN-EP-2022-002",
    doi = "10.1007/JHEP06(2022)005",
    journal = "JHEP",
    volume = "06",
    pages = "005",
    year = "2022"
}

@misc{HEPData,
  title = {ATLAS-EXOT-2019-23 HEPdata record},
  howpublished = {\url{https://www.hepdata.net/record/ins2043503}},
  note = {Accessed: 11 Jul 2023}
}

@misc{thomascode,
  title = {ATLAS-EXOT-2019-23 Recasting Code},
  howpublished = {\url{https://github.com/ThomasChehab/recastingCodes/tree/master/CalRatioDisplacedJet}},
  note = {Accessed: 27 Jul 2023}
}

@article{Gopalakrishna:2008dv,
    author = "Gopalakrishna, Shrihari and Jung, Sunghoon and Wells, James D.",
    title = "{Higgs boson decays to four fermions through an abelian hidden sector}",
    eprint = "0801.3456",
    archivePrefix = "arXiv",
    primaryClass = "hep-ph",
    reportNumber = "CERN-PH-TH-2008-11, MCTP-07-47, BNL-HET-08-2",
    doi = "10.1103/PhysRevD.78.055002",
    journal = "Phys. Rev. D",
    volume = "78",
    pages = "055002",
    year = "2008"
}

@article{Wells:2008xg,
    author = "Wells, James D.",
    editor = "Kane, Gordon and Pierce, Aaron",
    title = "{How to Find a Hidden World at the Large Hadron Collider}",
    eprint = "0803.1243",
    archivePrefix = "arXiv",
    primaryClass = "hep-ph",
    reportNumber = "MCTP-07-51, CERN-PH-TH-2008-47",
    pages = "283--298",
    month = "3",
    year = "2008"
}

@article{Curtin:2014cca,
    author = "Curtin, David and Essig, Rouven and Gori, Stefania and Shelton, Jessie",
    title = "{Illuminating Dark Photons with High-Energy Colliders}",
    eprint = "1412.0018",
    archivePrefix = "arXiv",
    primaryClass = "hep-ph",
    reportNumber = "YITP-SB-14-49",
    doi = "10.1007/JHEP02(2015)157",
    journal = "JHEP",
    volume = "02",
    pages = "157",
    year = "2015"
}

@article{Alloul:2013bka,
    author = "Alloul, Adam and Christensen, Neil D. and Degrande, C\'eline and Duhr, Claude and Fuks, Benjamin",
    title = "{FeynRules  2.0 - A complete toolbox for tree-level phenomenology}",
    eprint = "1310.1921",
    archivePrefix = "arXiv",
    primaryClass = "hep-ph",
    reportNumber = "CERN-PH-TH-2013-239, MCNET-13-14, IPPP-13-71, DCPT-13-142, PITT-PACC-1308",
    doi = "10.1016/j.cpc.2014.04.012",
    journal = "Comput. Phys. Commun.",
    volume = "185",
    pages = "2250--2300",
    year = "2014"
}

@article{Alwall:2014hca,
    author = "Alwall, J. and Frederix, R. and Frixione, S. and Hirschi, V. and Maltoni, F. and Mattelaer, O. and Shao, H. -S. and Stelzer, T. and Torrielli, P. and Zaro, M.",
    title = "{The automated computation of tree-level and next-to-leading order differential cross sections, and their matching to parton shower simulations}",
    eprint = "1405.0301",
    archivePrefix = "arXiv",
    primaryClass = "hep-ph",
    reportNumber = "CERN-PH-TH-2014-064, CP3-14-18, LPN14-066, MCNET-14-09, ZU-TH-14-14",
    doi = "10.1007/JHEP07(2014)079",
    journal = "JHEP",
    volume = "07",
    pages = "079",
    year = "2014"
}

@article{Bierlich:2022pfr,
    author = "Bierlich, Christian and others",
    title = "{A comprehensive guide to the physics and usage of PYTHIA 8.3}",
    eprint = "2203.11601",
    archivePrefix = "arXiv",
    primaryClass = "hep-ph",
    reportNumber = "LU-TP 22-16, MCNET-22-04, FERMILAB-PUB-22-227-SCD",
    doi = "10.21468/SciPostPhysCodeb.8",
    month = "3",
    year = "2022"
}

@article{Corpe:2024ntq,
    author = "Corpe, Louie and Goudelis, Andreas and Jeannot, Simon and Jeon, Si Hyun",
    title = "{Probing exotic long-lived particles from the prompt side using the CONTUR method}",
    eprint = "2407.18710",
    archivePrefix = "arXiv",
    primaryClass = "hep-ph",
    month = "7",
    year = "2024"
}

\end{document}